\documentclass[10pt]{article}

\usepackage{alltt}
\usepackage[pdftex]{graphicx}
\usepackage{amsmath,amsfonts}
\usepackage{scalerel}
\usepackage{subfigure}

\begin{document}

\title{Formal Scheduling Constraints for Time-Sensitive Networks}

\author
{Silviu S. Craciunas~~~~Ramon Serna Oliver~~~~Wilfried Steiner\\
\\
\normalsize{TTTech Computertechnik AG}\\
\normalsize{Sch{\"o}nbrunner Stra{\ss}e 7}\\
\normalsize{1040 Vienna, Austria}\\
\normalsize{\emph{forename.surname}\texttt{@tttech.com}}
}

\date{}

\maketitle
\thispagestyle{plain} 
\pagestyle{plain}

\begin{abstract}
In recent years, the $IEEE 802.1$ Time Sensitive Networking (TSN) task group has been 
active standardizing time-sensitive capabilities for \textit{Ethernet} networks ranging from 
distributed clock synchronization and time-based ingress policing to frame 
preemption, redundancy management, and scheduled traffic enhancements. In particular the scheduled 
traffic enhancements defined in IEEE~802.1Qbv together with the clock synchronization protocol 
open up the possibility to schedule communication in distributed networks providing 
real-time guarantees.

In this paper we formalize the necessary constraints for creating window-based IEEE~802.1Qbv Gate 
Control List schedules for Time-sensitive Networks (TSN). The resulting schedules allow a greater 
flexibility in terms of timing properties while still guaranteeing deterministic communication with 
bounded jitter and end-to-end latency.

\end{abstract}

\section{Introduction}
Deterministic real-time communications have long been a requirement in the aerospace 
domain~\cite{Orion, Kopetz93}. The strictness of certification programs and industry practices are 
only satisfied if sufficient proofs of evidence guarantee the deterministic behavior of  static 
configurations often deployed in production programs spanning over several decades. Recently, other 
fast-moving markets like automotive and industrial automation are increasingly joining the trend of 
deterministic networking albeit being less open to accept any detriment to generalized networking 
features, like high communication speeds with near-to-full bandwidth utilization, availability of 
off-the-shelf components, or dynamic cluster reconfigurations.

In recent years, the $IEEE 802.1$ Time Sensitive Networking (TSN) task group~\cite{TSN} has been 
active standardizing time-sensitive capabilities for \textit{Ethernet} networks ranging from 
distributed clock synchronization~\cite{802.1AS-rev} and time-based ingress policing~\cite{802.1Qci} 
to frame preemption~\cite{802.1Qbu}, redundancy management~\cite{802.1CB}, and scheduled traffic 
enhancements~\cite{802.1Qbv}.

Out of the many protocols being presented, two of them are key to achieving compositional 
real-time guarantees:
\begin{itemize}
	\item IEEE 802.1ASrev~\cite{802.1AS-rev} defines a time-synchronization protocol enabling a 
	global clock reference with basic fault-tolerance mechanisms. 
   \item IEEE 802.1Qbv~\cite{802.1Qbv} specifies the time-aware shaper functionality implementing 
   the time-triggered paradigm~\cite{kopetz03} at the egress port of communicating nodes.
\end{itemize}

A time-aware shaper is essentially a gate mechanism sitting at the egress side of each priority 
queue dynamically enabling or disabling the selection of frames from the respective queue based on 
a predefined periodic schedule referred as \textit{Gate Control List}. In a distributed network 
each egress port is timely set at run-time according to its own configured gate control list 
executed synchronously according to the global notion of time. 

In the combination of these two key features lay the foundations for the synthesis of schedules 
driving the communication in a distributed network with determinism and end-to-end real-time 
guarantees.

Previous work~\cite{CraciunasRTNS16} formally defined necessary constraints to compute deterministic 
schedules that could be mapped to TSN-compliant multi-hop switched networks providing jitter-free 
transmission and deterministic end-to-end latency guarantees for strictly-periodic scheduled frames. 
However, such stringent requirements on jitter and latency came at a high cost. On one hand, fully 
deterministic communication constraints restrict the solution space for valid schedules due to the 
isolation of streams in the time domain. On the other hand, the focus was given to finding exact 
timing for each transmitted frame, which was then mapped on a second step into a gate control list 
reproducing the expected behavior. This made it difficult to optimize and tailor the output to 
device-specific properties, like the length of the resulting gate control list or the minimum 
distance between consecutive open and close gate operations. 

\sloppypar
In this paper we elaborate the necessary constraints to synthesize $IEEE 802.1Qbv$ deterministic
schedules with a focus on the gate operations (open/close) and a relaxed timing model allowing 
communication with bounded jitter, yet guaranteeing determinism.

We introduce the network and traffic model in Section~\ref{sec:model} and formulate the 
scheduling constraints for IEEE 802.1Qbv time-gates as well as for the defined communication streams 
(Section~\ref{sec:constraints}) for the case in which all communication streams share the same period. We discuss in~\ref{sec:multiperiod} the extension to multiple periods followed by a description of an SMT-based synthesis algorithm 
implementing the previously defined constraints in Section~\ref{sec:scheduling}. 

\section{System Model}
\label{sec:model}

\begin{figure}[h!]  
\begin{center}
     \includegraphics[width=8cm]{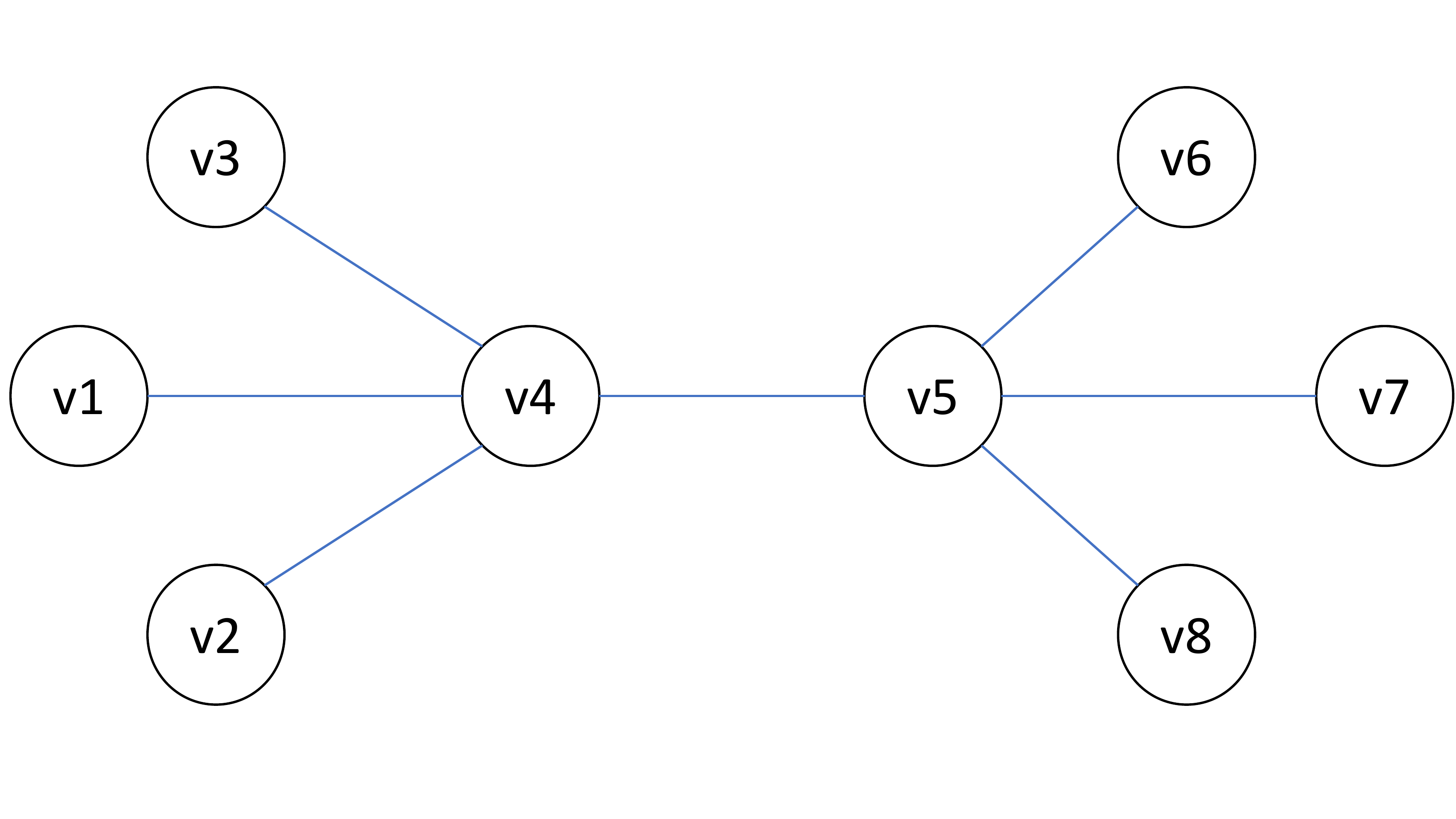} \end{center}
     \caption{Example network}
	\label{fig:example-network}
\end{figure}

In this paper we model networks as a graph $\mathcal{G}=\{\mathcal{V}, \mathcal{E}\}$, 
where $\mathcal{V}$ is a set of vertices, $\mathcal{E}$ is a set of non-directed edges as well as 
directed edges connecting vertices to each other. Each undirected edge $(v_i,v_j) \in \mathcal{E}$ 
between two vertices $v_i, v_j \in \mathcal{V}$ defines two directed edges $[v_i,v_j], [v_j,v_i] \in 
\mathcal{E}$ between the two vertices, where the first vertex in the pair description defines the 
source vertex and the second vertex defines the destination vertex. An example graph $\mathcal{G}$ 
with eight vertices and seven undirected edges resulting in fourteen directed edges is depicted in 
Figure~ \ref{fig:example-network}.

Vertices may source, forward, and/or receive messages. End-nodes can be the source or the destination of streams, while network switches are the intermediary nodes forwarding those messages. 
A communication requirement is introduced through the concept of stream. A stream (or flow) is a 
periodic multicast data transmission from one talker (the sender) to one or multiple listeners (the receivers).
Without loss of generality, we restrict the number of receivers to one (unicast) and the message size to one
single frame, noting that extending the model to the general case is a trivial step~\cite{Craciunas15}.
We denote the set of streams in the network with $\mathcal{S}$. 
Similar to~\cite{steiner10}, a stream $s_i \in \mathcal{S}$ from talker node $v_1$ to listener node 
$v_n$ routed through the intermediary nodes (i.e. switches) $v_2, v_3 \ldots, v_{n-1}$ 
is expressed as $$s_i = [[v_1,v_2], \ldots, [v_{n-1},v_n]].$$ 
We assume that for each stream the sender and receiver vertices $v_1, v_n$, as well as the 
routed communication path that connect the sender and receiver vertices are known and given.
 
A stream is defined by the tuple $\langle s_i.\mathit{e2e}, \langle s_i.\mathit{jitter}, s_i.\mathit{size}, s_i.\mathit{period} \rangle,$ 
denoting the maximum allowed end-to-end latency, the maximum allowed jitter, the data size in bytes, and the period of the flow, 
respectively. For the main part of the constraints we assume that all streams in the system share the same period but discuss the extension to multiple periods in Section~\ref{sec:multiperiod}. 

The instance of a stream $s_i \in \mathcal{S}$ routed through link $[v_a, v_b] \in 
\mathcal{E}$ is defined by a frame $f_i^{[v_a,v_b]} \in \mathcal{F}^{[v_a,v_b]}$, where 
$\mathcal{F}^{[v_a,v_b]}$ is the set of all frames that are to be scheduled on link $[v_a, v_b]$.
Each such periodic frame is characterized by a frame length $f_i^{[v_a,v_b]}.\mathit{L}$ and a 
frame period $f_i^{[v_a,v_b]}.\mathit{T}$. The period of the frame is equal to the period of 
the stream while the length of the frame is calculated based on the data size of the stream and 
the link speed.

Each vertex implements at least one queue $q$ for each directed edge that it sources. For example, 
$v_4$ implements a queue $q^{[v_4,v_5]}_i$ for the edge $[v_4,v_5]$.  Thus, all scheduled frames to be 
communicated from one vertex to another one will traverse the same queue. In this paper we will assume one 
single queue for scheduled traffic.

\paragraph*{Time Aware Shaper}

IEEE 802.1Qbv~\cite{802.1Qbv} defines a specific shaping mechanism on how frames are selected for 
transmission on egress. In particular it defines a \emph{gate} for each priority queue, which at any time 
is in one of the two states \emph{open} or \emph{close}. When the gate of a respective queue is in the 
open state, frames may be selected for transmission on the directed edge associated with the queue 
in first-in first-out (FIFO) order. In case the gate of a respective queue is in the close state, 
frames from this queue are not selected. A priority selection is then applied among all opened queues.
The state changes are predefined with respect to a global time via the Gate Control List, 
which is cyclically executed at runtime. 
The synchronized time (global timebase) can be established by an appropriate 
synchronization protocol as IEEE 802.1AS~\cite{802.1AS-rev}. Vertexes  
continually check whether a state change for one of its gates is scheduled in the gate control list
and apply it when due. 

Encoded in the gate control list is an ordered list of transmission windows on a timeline, i.e., 
windows during which a gate is in the open state. Each window $w^{[v_a,v_b]}_k$ is defined by a 
left boundary $w^{[v_a,v_b]}_k.\mathit{open}$ and a right 
boundary $w^{[v_a,v_b]}_k.\mathit{close}$. As we will see later, the maximum number of windows  
$\mathcal{W}_\mathit{max}^{[v_a,v_b]}$ per edge derived
from the maximum length of the gate control list will be an essential parameter in the schedule 
synthesis. 
We define a boolean array $w^{[v_a,v_b]}_k.\epsilon(f_i^{[v_a,v_b]})$ for 
each window $w^{[v_a,v_b]}_k$ that describes whether frame $f_i^{[v_a,v_b]} \in \mathcal{F}^{[v_a,v_b]}$
is assigned to the window or not.

\section{Formal Scheduling Constraints}
\label{sec:constraints}
In this section we enumerate the constraints for the gate open and close operations as well as
the frame to window assignment variables leading to the computation of a gate control list such 
that the correct temporal behaviour on the scheduled streams is guaranteed.

\paragraph{Well-defined windows constraints}

We initially introduce for each link the constraints restricting the open and close events of each 
window defined on that link to be within the hyperperiod of all streams. 
The open event for each window on a link has to be greater than or equal to $0$, hence we have the 
constraint
\begin{align}
&\forall [v_a,v_b] \in \mathcal{E}, \forall k \in \{1,\dots,\mathcal{W}_\mathit{max}^{[v_a,v_b]}\}: \nonumber\\ 
&w^{[v_a,v_b]}_k.\mathit{open} \geq 0 \label{c2} \\
&w^{[v_a,v_b]}_k.\mathit{close} < \mathit{hp}^{[v_a,v_b]}\nonumber
\end{align}
where $\mathit{hp}^{[v_a,v_b]} \stackrel{def}{=} lcm(\{f_i^{[v_a,v_b]}.\mathit{T}\mid f_i^{[v_a,v_b]} 
\in \mathcal{F}^{[v_a,v_b]}\})$ is the hyperperiod of all frames that are routed through $[v_a,v_b]$. 

Additionally, well-defined windows must ensure the order between the gate open and close events.
Hence we have 
\begin{align*}
&\forall [v_a,v_b] \in \mathcal{E}, \forall k \in \{1,\dots,\mathcal{W}_\mathit{max}^{[v_a,v_b]}\}: \\ 
&w^{[v_a,v_b]}_k.\mathit{open} \leq w^{[v_a,v_b]}_k.\mathit{close} \nonumber
\end{align*}

We restrict the variables defining the assignment of frames to windows to get a boolean value, i.e., they 
can be either $0$ or $1$. 
\begin{align*}
&\forall [v_a,v_b] \in \mathcal{E}, \forall k \in \{1,\dots,\mathcal{W}_\mathit{max}^{[v_a,v_b]}\}, 
\forall f_i^{[v_a,v_b]} \in \mathcal{F}^{[v_a,v_b]}: \\
&w^{[v_a,v_b]}_k.\epsilon(f_i^{[v_a,v_b]}) \in \{0,1\}   \nonumber
\end{align*}

\paragraph{Ordered windows constraints}
An essential constraint for determinism is that no two frames transmitted on the same egress link 
overlap in the time domain. In~\cite{CraciunasRTNS16} we 
constructed a similar constraint by not allowing frames to be transmitted at the same time. Since in this 
paper we are considering the scheduling of windows rather than frames we consider two possible formulations.

One possibility, similar to~\cite{CraciunasRTNS16}, is to not allow windows on the same link to overlap, defined as

\begin{align*}
&\forall [v_a,v_b] \in \mathcal{E}, \forall k,l \in \{1,\dots,\mathcal{W}_\mathit{max}^{[v_a,v_b]}\}, k \neq l: \\
&(w^{[v_a,v_b]}_k.\mathit{close} \leq w^{[v_a,v_b]}_{l}.\mathit{open})  \vee \\ \nonumber
&(w^{[v_a,v_b]}_l.\mathit{close} \leq w^{[v_a,v_b]}_{k}.\mathit{open}) \nonumber
\end{align*}

The drawback with this option is that it results in a large number of assertions with a disjunction operator,
which proved to be computationally intensive.

However, since there is no predefined assignment of frames to windows we can fix the order of windows on a link and
relate their respective open and close events sequentially. Therefore, we prefer the following alternative formulation,
which is equivalent to the above but significantly less resource intensive.
\begin{align*}
&\forall [v_a,v_b] \in \mathcal{E}, \forall k \in \{1,\dots,\mathcal{W}_\mathit{max}^{[v_a,v_b]} - 1\}: \\
&w^{[v_a,v_b]}_k.\mathit{close} \leq w^{[v_a,v_b]}_{k+1}.\mathit{open}   \nonumber
\end{align*}

Additionally, the latter formulation allows reducing further the number of assertions in constraints~\eqref{c2} by restricting the bound on the open event to the first window and, respectively, the close event to the last window on each link since all others are already bounded due to the imposed sequential order. Hence, constraint~\eqref{c2} can be reduced to 
\begin{align*}
&\forall [v_a,v_b] \in \mathcal{E} : \\
&w^{[v_a,v_b]}_1.\mathit{open} \geq 0 \nonumber \\
&w^{[v_a,v_b]}_{\mathcal{W}_\mathit{max}^{[v_a,v_b]}}.\mathit{close} < \mathit{hp}^{[v_a,v_b]} \nonumber
\end{align*}

\paragraph{Frame-to-window assignment constraints}
Each frame routed through a link has to be assigned to one and only one window defined for that link.
Since the frame assignment variables in the array $\epsilon$ can either take the value $0$ or $1$, we formulate that the 
sum of all window assignment variables for that particular frame to be exactly $1$. 
\begin{align*}
&\forall [v_a,v_b] \in \mathcal{E}, \forall f_i^{[v_a,v_b]} \in \mathcal{F}^{[v_a,v_b]}: \\
&\sum_{k = 1}^ {\mathcal{W}_\mathit{max}^{[v_a,v_b]}} w^{[v_a,v_b]}_k.\epsilon(f_i^{[v_a,v_b]}) = 1 \nonumber
\end{align*}

\paragraph{Window size constraints}
We define the close event of each window sufficient to allow the transmission of all the frames assigned to it.
Therefore, we calculate the size of each window based on the frame assignment variables and the respective size of the frames.
\begin{align*}
&\forall [v_a,v_b] \in \mathcal{E}, \forall k \in \{1,\dots,\mathcal{W}_\mathit{max}^{[v_a,v_b]}\}: \\
&w^{[v_a,v_b]}_k.\mathit{size} = \sum_{\scaleto{f_i^{[v_a,v_b]} \in \mathcal{F}^{[v_a,v_b]}}{8pt}} 
w^{[v_a,v_b]}_k.\epsilon(f_i^{[v_a,v_b]}) \times f_i^{[v_a,v_b]}.\mathit{L} \nonumber
\end{align*}

Using the formula for the window size constructed before, we introduce the constraint that the time 
interval between open and close events of a window has to be equal to the window size.
\begin{align*}
&\forall [v_a,v_b] \in \mathcal{E}, \forall k \in \{1,\dots,\mathcal{W}_\mathit{max}^{[v_a,v_b]}\}: \\
&w^{[v_a,v_b]}_k.\mathit{close} = w^{[v_a,v_b]}_{k}.\mathit{open} + w^{[v_a,v_b]}_{k}.\mathit{size} \nonumber
\end{align*}

\paragraph{Stream constraints}
The stream constraints describe the sequential nature of a communication from a talker (sender) node 
to a listener (receiver) node. The generic condition is that frames belonging to the same stream 
have to be scheduled sequentially on the time-line along the routed communication path. In other 
words, the propagation of frames of a stream must follow the sequential order along the computed 
route of the stream.

For every stream $s_i \in \mathcal{S}$ routed through $v_1, ..., v_n$ we construct the following 
formula
\begin{align*}
&\forall [v_j, v_{j+1}] \in \mathcal{E}, j \in \{1, \dots, n-2\}, \\
&\forall k \in \{1,\dots,\mathcal{W}_\mathit{max}^{[v_j, v_{j+1}]}\}, \forall l \in \{1,\dots,\mathcal{W}_\mathit{max}^{[v_{j+1}, v_{j+2}]}\}: \nonumber\\
& \chi_{i,k,l} =  w^{[v_j, v_{j+1}]}_{k}.\epsilon(f_i^{[v_j, v_{j+1}]}) \times w^{[v_{j+1}, v_{j+2}]}_{l}.\epsilon(f_i^{[v_{j+1}, v_{j+2}]})  \nonumber\\
& \chi_{i,k,l} \times (w^{[v_j, v_{j+1}]}_{k}.\mathit{close} + \delta) \leq \chi_{i,k,l} \times w^{[v_{j+1}, v_{j+2}]}_{l}.\mathit{open}\nonumber
\end{align*}

where $\delta$ represents the worst-case difference between the 
local clocks of any two synchronized (e.g. via the IEEE 802.1AS~\cite{802.1AS-rev} 
time-synchronization protocol) vertexes. Hence, we also consider, similar to~\cite{Zhang14}, the 
synchronization jitter which is a global constant and describes the maximum difference between the 
local clocks of any two nodes in the network.

\paragraph{Stream isolation constraints}
As described in~\cite{CraciunasRTNS16}, a network may experience frame loss or variations in 
periodic payload size during the runtime of the system. Since the IEEE 802.1Qbv~\cite{802.1Qbv} 
specification controls the opening and closing of the timed gates of a queue and not the sending and 
receiving of individual frames, we need to ensure that the state of the queue is deterministic.
We refer the reader to~\cite{CraciunasRTNS16} for a more detailed description of the isolation 
problem in TSN networks. 

Consider the case in which two streams $s_i$ and $s_j$ are received on different links, $[v_x,v_a]$ 
and $[v_y,v_a]$, respectively, on device $v_a$ and are both sent on the same egress port on link 
$[v_a,v_b]$. In order to maintain their respective window assignments on the egress port even in the 
case of frame loss or concurrent arrival we need to either assign the respective frames of the two 
streams to the same window or to isolate them in the time domain. The isolation in the time domain is done by 
restricting that once the frame of one stream has entered the device, the other stream cannot enter 
it until the first frame has left the egress queue (c.f.~\cite{CraciunasRTNS16}). If there is more 
than one queue available for scheduled traffic we can also isolate the two streams in different 
queues. Since we restrict the TSN configuration for this paper to $1$ queue for scheduled traffic we 
construct the isolation constraint without the option to assign streams to different queues noting 
that the extension to multiple queues is straightforward.

Hence, we have the stream isolation condition:
\begin{align}
&\forall [v_a, v_b] \in \mathcal{E}, \forall f_i^{[v_a, v_b]}, f_j^{[v_a, v_b]} \in \mathcal{S}, i\neq j, \nonumber\\
&\forall k \in \{1,\dots,\mathcal{W}_\mathit{max}^{[v_a, v_b]}\}, \nonumber \\
&\forall l \in \{1,\dots,\mathcal{W}_\mathit{max}^{[v_x, v_a]}\}, \nonumber \\
&\forall m \in \{1,\dots,\mathcal{W}_\mathit{max}^{[v_y, v_a]}\}: \nonumber \\
&\Bigl(w^{[v_a,v_b]}_{k}.\mathit{close} \times w^{[v_a,v_b]}_{k}.\epsilon(f_i^{[v_a, v_b]}) \le \nonumber \\
&w^{[v_y,v_a]}_{m}.\mathit{open} \times w^{[v_y,v_a]}_{m}.\epsilon(f_j^{[v_y, v_a]}) \vee \\
&w^{[v_a,v_b]}_{k}.\mathit{close} \times w^{[v_a,v_b]}_{k}.\epsilon(f_j^{[v_a, v_b]}) \le \nonumber \\
&w^{[v_x,v_a]}_{l}.\mathit{open} \times w^{[v_x,v_a]}_{l}.\epsilon(f_i^{[v_x, v_a]})\Bigr) \vee \nonumber \\
&\Bigl(w^{[v_a,v_b]}_{k}.\epsilon(f_j^{[v_a, v_b]}) = w^{[v_a,v_b]}_{k}.\epsilon(f_i^{[v_a, v_b]})\Bigr) \nonumber
\end{align}

\paragraph{Stream End-to-end latency constraints}
The end-to-end deadline constraint states that the difference between the receiving of a stream on 
the listener and the sending of the stream from the respective talker has to be smaller than or equal 
to the given maximum end-to-end latency. However, since the frame to window assignment is not known a-priori, 
we construct the formula using the frame assignment variables in combination with the open and close events.

Let $v_1^i$ and $v_n^i$ be the talker and listener nodes of stream $s_i \in \mathcal{S}$, respectively, 
and $f_i^{[v_1^i, v_2^i]}$ and $f_i^{[v_{n-1}^i, v_n^i]}$ the first and last frames of the stream from/to those 
nodes. We define the end-to-end latency constraint as
\begin{align*}
&\forall s_i \in \mathcal{S}: \\
&\sum_{l = 1}^{\scaleto{\mathcal{W}_\mathit{max}^{[v_{n-1}^i, v_n^i]}}{8pt}} w^{[v_{n-1}^i, v_n^i]}_{l}.\epsilon(f_i^{[v_{n-1}^i, v_n^i]}) \times w^{[v_{n-1}^i, v_n^i]}_{l}.\mathit{close} - \nonumber\\
&\sum_{k = 1}^{\scaleto{\mathcal{W}_\mathit{max}^{[v_1^i, v_2^i]}}{8pt}} w^{[v_1^i, v_2^i]}_{k}.\epsilon(f_i^{[v_1^i, v_2^i]}) \times w^{[v_1^i, v_2^i]}_{k}.\mathit{open} \leq \nonumber \\
&s_i.\mathit{e2e} - f_i^{[v_{n-1}^i, v_n^i]}.\mathit{L} - \delta. \nonumber
\end{align*}

Note that we also include the precision $\delta$ in the constraint since the 
local clocks of the sender and receiver nodes may show a synchronization error with respect to each 
other.

\paragraph{Stream jitter constraints}
Real-time communication also may require constraints on the jitter that a stream experiences. We 
base our jitter constraint on the observation that within the network, the jitter of individual 
frames of a stream is not relevant. The jitter becomes relevant on the receiver side when it has to 
be processed by a listener task. Hence, we only constrain the jitter of a stream for the receiver, 
i.e., the jitter constrain only applies to the sending of a stream on the last hop before the 
listener node. 

As above, let $v_n^i$ be the listener node of stream $s_i \in \mathcal{S}$ with $f_i^{[v_{n-1}^i, 
v_n^i]}$ the last frames of the stream to the listener node.
\begin{align*}
&\forall s_i \in \mathcal{S}, \\
&\psi(s_i) = \sum_{l = 1}^{\scaleto{\mathcal{W}_\mathit{max}^{[v_{n-1}^i, v_n^i]}}{8pt}} 
w^{[v_{n-1}^i, v_n^i]}_{l}.\epsilon(f_i^{[v_{n-1}^i, v_n^i]}) \times w^{[v_{n-1}^i, v_n^i]}_{l}.\mathit{size}: \nonumber\\
&\psi(s_i) \leq s_i.\mathit{jitter} + f_i^{[v_{n-1}^i, v_n^i]}.\mathit{L} \nonumber
\end{align*}

Please note that, if the jitter is also important within the network, i.e. between individual nodes 
along the route of the stream, the constraint defined above can be readily applied for each of those 
nodes. 

\section{Multiple periods}
\label{sec:multiperiod}
So far we have assumed that all streams share the same period. Communication in system deployments do not always appear with a normalized period. Instead, streams are sourced at multiple rates which result in a hypercycle defining the length of the schedule tables to be at least the \textit{least common multiple} of all periods involved. In~\cite{CraciunasRTNS16} the communication model guaranteed minimal jitter by imposing a strictly periodic constraint between frames belonging to alternative scheduled instances of the same stream. In this paper, however, we introduce a relaxed model allowing bounded jitter between periodic instances of a stream.

The assignment of frames, and as a consequence the length of each window, is a result of the scheduler.
Furthermore, frames routed through the same link may have different periods and each instances of a frame 
may be assigned to different windows. Hence, in order to correctly constrain the closing bound for each window we need to consider the periodic repetition of stream instances within the hyperperiod. For this, we refine the concept of stream instance $s_i \in \mathit{S}$ to reflect the number of streams instance repetitions $s_{i,j}$, where $0 \leq j \leq \frac{\mathit{hp}^{[v_a,v_b]}}{s_i.T} - 1$ corresponding to the instantiation of the stream $s_i$ for each period interval completing the schedule hyperperiod. Therefore, each original stream $s_i$ will result in an effective number of streams equal to the integer division between the hyperperiod and the stream period. Furthermore, we define the subset of frame instances $f_{i,j}^{[v_a,v_b]} \in \mathcal{F}_j^{[v_a,v_b]}$ as the  frames resulting from the $j^{th}$ repetition of $s_i$ routed through $[v_a,v_b]$, sorted in ascending ordered by the period instance.

Therefore, for each stream $s_i$ routed through $[v_a,v_b]$ we construct the following constraint:

\begin{align*}
&\forall j \in \left[0, \frac{\mathit{hp}^{[v_a,v_b]}}{s_i.T} - 1 \right], 
\forall k \in \{1,\dots,\mathcal{W}_\mathit{max}^{[v_a, v_b]}\}: \\
&\beta^{[v_a,v_b]}_{k,i,j} = w^{[v_a, v_b]}_{k}.\epsilon(f_{i,j}^{[v_a, v_b]}) \nonumber\\
&\beta^{[v_a,v_b]}_{k,i,j} \times w^{[v_a, v_b]}_{k}.\mathit{open} \geq \beta^{[v_a,v_b]}_{k,i,j} \times j \times f_{i,j}^{[v_a, v_b]}.T \nonumber\\
&\beta^{[v_a,v_b]}_{k,i,j} \times w^{[v_a, v_b]}_{k}.\mathit{close} \leq \beta^{[v_a,v_b]}_{k,i,j} \times (j + 1) \times f_{i,j}^{[v_a, v_b]}.T \nonumber\\
\end{align*}

Note that this constraint would be sufficient to bound the open and close events of all windows for which at least one frame is assigned. For completion, we leave the above formulation (see first constraint \textit{well-defined windows constraints}) setting bounds for all open and close window events, including those without any assigned frame, to be in the range $[0..hp^{[v_a,v_b]})$.

Having multiple periods in the system also has an effect on the jitter and end-to-end latency 
constraint. The end-to-end latency constraint is easy to reformulate since it has to consider that 
the difference between all instances of sending and receiving frames of a stream has to 
conform to the given maximum latency. The jitter constraint is more difficult to reformulate since 
it requires the computation of the jitter between the earliest window open event within a period 
and the latest window close event within a period for all frame instance assignments. We leave the formulation of this constraint for future work. 

\section{SMT-Based Schedule Synthesis}
\label{sec:scheduling}
Satisfiability Modulo Theories (SMT) determine the satisfiability of first-order 
logical formulas for a specific background theory like linear integer arithmetic 
($\mathcal{LA}(\mathbb{Z}$)) or bit-vectors ($\mathcal{BV}$)~\cite{barrett09, sebastiani07} while 
also providing a \textit{model} (in case of satisfiability) which represents one solution for 
the given SMT context. In addition, Optimization Modulo Theories (SMT)~\cite{st_cav15, Bjorner2015}
can provide optimal solutions with respect to certain objectives. 

The aim of our scheduling algorithm for IEEE 802.1Qbv is hence to find solutions for the window open and 
close variables as well as for the frame-to-window assignment variables such that the correctness 
constraints defined in Section~\ref{sec:constraints} are fulfilled. While in previous work the 
background theory used was linear integer arithmetic ($\mathcal{LIA}$), in our case we 
have to use non-linear integer arithmetic ($\mathcal{NIA}$). While nonlinear integer arithmetic has 
been shown to be undecidable, solvers use methods like bit-blasting to solve problems with limited 
range integer variables in $\mathcal{NIA}$~\cite{6821146}.

A typical optimization objective found in the industrial domain is to minimizing the end-to-end 
latencies of streams. This optimization objective is easily expressed based on the end-to-end 
constraint defined in Section~\ref{sec:constraints} and has been discussed in related work 
(cf.~\cite{Craciunas15}). 

Since there is a trade-off between the number of windows used per egress port and the jitter that a 
stream experiences, another optimization objective is to minimizes the 
receiving jitter for streams. This objective can be either expressed as a sum over all streams, 
minimizing the accrued jitter in the network or as a collection of individual objectives which 
result in local minima for some some streams. 

\section{Conclusion}
\label{sec:conclusion}
We have formalized the necessary scheduling conditions for creating window-based IEEE~802.1Qbv Gate 
Control List schedules for Time-sensitive Networks (TSN) in which communication streams have 
real-time requirements in terms of bounded jitter and end-to-end latency.

\bibliographystyle{IEEEtran}
\bibliography{bibliography}

\end{document}